\documentclass[a4paper, twocolumn]{article}

\usepackage{cite}
\usepackage{amsmath,amssymb,amsfonts}
\usepackage{graphicx}
\usepackage{listings}
\usepackage{textcomp}
\usepackage{xcolor}
\def\BibTeX{{\rm B\kern-.05em{\sc i\kern-.025em b}\kern-.08em
    T\kern-.1667em\lower.7ex\hbox{E}\kern-.125emX}}

\begin{document}

\title{Proof of Authenticity of General IoT Information with Tamper-Evident Sensors and Blockchain
\thanks{This manuscript has been accepted for a full paper presentation at IEEE Region 10 Humanitarian Technology Conference 2025.
This research has been supported by JSPS KAKENHI Grant Number JP24H00691.}
}

\makeatletter
\newcommand{\linebreakand}{%
  \end{@IEEEauthorhalign}
  \hfill\mbox{}\par
  \mbox{}\hfill\begin{@IEEEauthorhalign}
}
\makeatother

\author{Kenji Saito\\
Graduate School of Business and Finance\\
Waseda University
}

\date{}

\maketitle

\begin{abstract}
Sensor data in IoT (Internet of Things) systems is vulnerable to tampering or falsification when transmitted through untrusted services.
This is critical because such data increasingly underpins real-world decisions in domains such as logistics, healthcare, and other critical infrastructure.
We propose a general method for secure sensor-data logging in which tamper-evident devices periodically sign readouts, link data using redundant hash chains, and submit cryptographic evidence to a blockchain-based service via Merkle trees to ensure verifiability even under data loss.
Our approach enables reliable and cost-effective validation of sensor data across diverse IoT systems, including disaster response and other humanitarian applications, without relying on the integrity of intermediate systems.
\end{abstract}

\begin{description}
\item[Keywords:]
Internet of Things (IoT), Blockchain, Sensor data authentication, Tamper-evident systems, Secure data logging, Humanitarian technology, Disaster response systems
\end{description}

\section{Introduction}
Sensor-generated data plays a vital role in emerging applications such as autonomous driving, surveillance, smart infrastructure, and healthcare.
However, the trustworthiness of this data is often compromised when routed through cloud-based IoT services or third-party intermediaries.
Inspired by our prior work\cite{Watanabe2021:RFID} on RFID authenticity with blockchain, we generalize the concept to support various types of sensors and actuators even under data loss.
Our goal is to ensure that users can verify sensor data provenance and integrity even in unstable, hostile, or opaque environments.

This challenge becomes especially critical in humanitarian scenarios such as disaster response, where network infrastructure may be unstable or damaged, and data from environmental sensors, rescue robots, or temporary medical stations must still be trusted for life-critical decisions.
In such settings, the ability to independently verify sensor data---even in the presence of partial losses or untrusted intermediaries---becomes vital for operational safety and accountability.

\section{Background}
To follow this paper, the readers should be familiar with the following concepts: cryptographic hash functions and (message) digests as their output, digital signatures, blockchain\cite{Nakamoto2008:Bitcoin}, and Merkle trees\cite{Merkle1988:Tree}.
This section provides brief background information on blockchain and proofs using Merkle trees, which may not be familiar to the readers.

Our prior work in RFID security \cite{Watanabe2021:RFID} and verifiable selective disclosure \cite{Saito2021:SelectiveDisclosure} also lays the foundation for our proposal.

\subsection{Design Goals of Blockchain}
The first design document describing Bitcoin\cite{Nakamoto2008:Bitcoin} began with the problem of transferring funds through a financial institution as a trusted third party.
Such a third party could, in principle, censor transfers and deny the transfer of funds in the form of freezing accounts, for example.
The goal of Bitcoin's design, therefore, is a remittance system that excludes such third parties and cannot be censored by any person.
This concept can be extended to the recording of events in general, as was later realized by the Ethereum\cite{Buterin2013:Ethereum} blockchain.
The generalized goal is the realization that ``No one can stop anyone from recording an event.''

This requirement can be broken down into the following four properties:
\begin{enumerate}
\item Self-sovereignty --- Users can participate in the system only at their own will without requiring permissions from others, and can direct the recordings of events.
\item Censorship resistance in the narrow sense --- Recordings directed by the user cannot be stopped by the will of anyone else.
\item Fault Tolerance --- Recordings directed by the user will not be stopped by a system breakdown or failure.
\item Tamper-resistant --- Records cannot be deleted or changed later, and records that were not there in the past cannot be falsified.
\end{enumerate}

If all of the above are satisfied, it will make a record censorship-resistant in the broadest sense (i.e., the record cannot be denied by anyone or anything).
Challenges exist as to whether the existing blockchains are always able to meet these properties\cite{Saito2016:Blockchain}.
However, such issues are independent of the discussion in this paper, and henceforth we will assume that blockchains to be used in our proposal satisfy these properties.

\subsection{Merkle Proof for Evidences}
\begin{figure*}
\begin{center}
\includegraphics[scale=0.6]{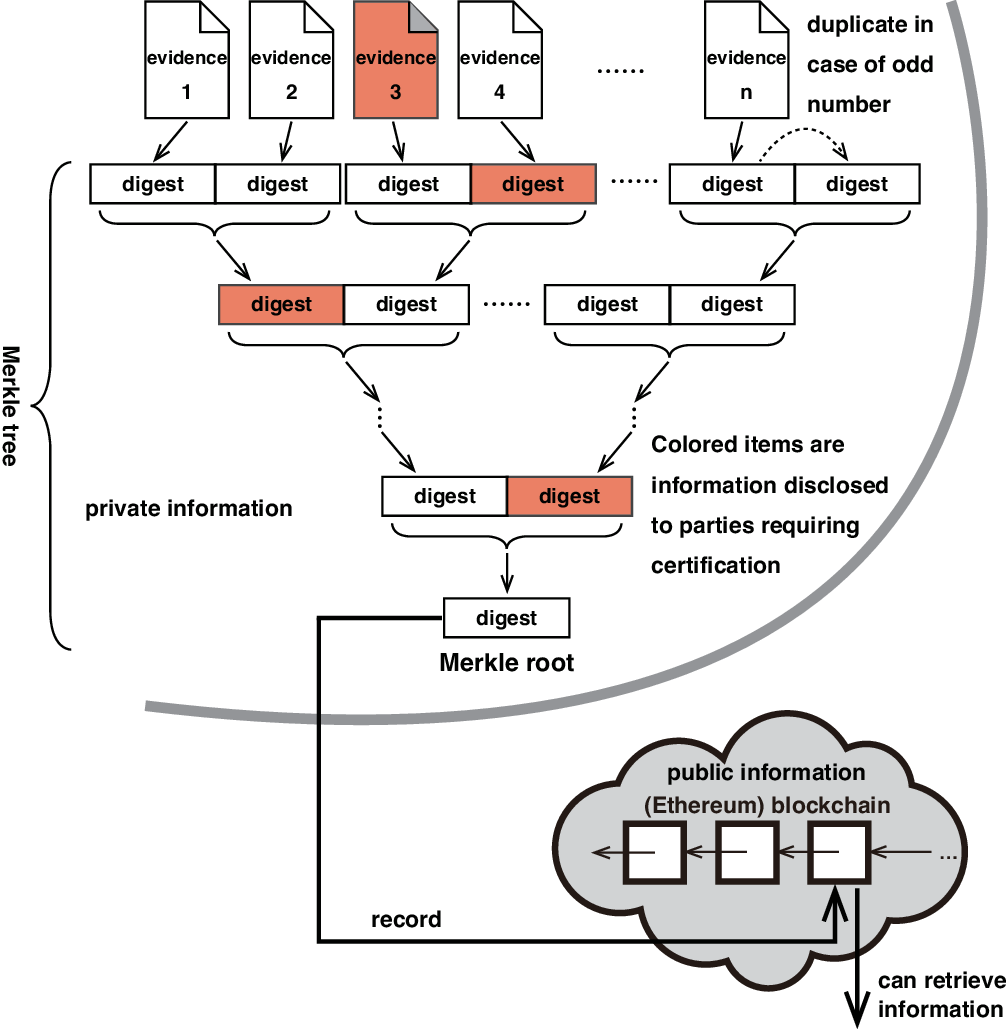}
{\footnotesize
\begin{itemize}
\item Starting with the digest of evidence 3, verifier will know the series of digests to be concatenated, so they can reproduce the calculations down to the Markle root, and confirm that the root matches the value recorded in the (Ethereum blockchain) smart contract.
\end{itemize}
}
\caption{Merkle proof for digest verification}\label{fig-merkle}
\end{center}
\end{figure*}

Existing blockchains such as Bitcoin and Ethereum bundle multiple records into blocks and link them using a hash chain, where each block contains the digest of the previous one.
These systems are designed so that any attempt to recreate or alter the chain must incur costs denominated in the blockchain's native cryptocurrency \cite{Iwamura2019:BitcoinMonetaryPolicy}\cite{Saito2023:ETHPOS}.
As the market value of these native tokens remains high, the structure effectively becomes resistant to censorship or tampering.
This makes the cost of writing tend to be high because a fee for writing must be paid in said cryptocurrency.
Therefore, if one wants to design a service that uses blockchain to record evidence, it is necessary to devise a way to lower the cost of recording.
One such device is Merkle proof (Fig.~\ref{fig-merkle}) utilizing a Merkle tree.

By using a Merkle tree, which is a type of hash tree, it becomes possible to verify the existence and authenticity of any individual piece of evidence by writing only a representative value (the Merkle root) to the blockchain.

\begin{figure*}
\begin{center}
{\scriptsize
\begin{lstlisting}[frame=single]
contract BBcAnchor {
  mapping (uint256 => uint) public _digests;
  constructor () public {
  }
  function getStored(uint256 digest) public view
                            returns (uint block_no) {
    return (_digests[digest]);
  }
  function isStored(uint256 digest) public view
                            returns (bool isStored) {
    return (_digests[digest] > 0);
  }
  function store(uint256 digest) public
                            returns (bool isAlreadyStored) {
    bool isRes = _digests[digest] > 0;
    if (!isRes) {
      _digests[digest] = block.number;
    }
    return (isRes);
  }
}
\end{lstlisting}
}
\caption{Smart contract for digest storage}\label{fig-anchor}
\end{center}
\end{figure*}

When writing the Merkle root to a blockchain, a smart contract can be written and deployed for this purpose---such as on the Ethereum blockchain. Fig.~\ref{fig-anchor} shows an excerpt from the actual smart contract code used in our Beyond Blockchain (BBc)\footnote{
https://github.com/beyond-blockchain
} project.
This contract saves the current block number for a stored digest.

\subsection{Related Prior Work}
This work builds on our previous studies in two directions: the authenticity of RFID-based logistics data and lightweight selective disclosure of verifiable documents using blockchain.

In our earlier work on RFID-based logistics systems\cite{Watanabe2021:RFID}, we proposed an architecture in which digitally signing, tamper-evident RFID readers transmit tag readouts to a logistics service and corresponding cryptographic evidence to a blockchain service in an atomic action.
The evidence is aggregated into a Merkle tree and periodically committed to the Ethereum blockchain, enabling verification of data integrity even if private keys are compromised or certificates expire.
A working prototype demonstrated the feasibility and cost-effectiveness of the approach.

In a separate study on selective disclosure\cite{Saito2021:SelectiveDisclosure}, we developed a scheme for documents where parts can be hidden or revealed in a verifiable way, using hash-based commitments and partial digital signatures.
By representing documents as XML structures and aggregating multiple such commitments into a single digest, we enabled scalable proof of existence and authenticity on a blockchain without exposing unnecessary data.

These two lines of work provided the foundational mechanisms---Merkle tree–based evidence structuring, blockchain-backed verification, and partial data disclosure---that are generalized and unified in the present study to support a wider range of IoT scenarios involving sensors and actuators.

\section{Problem Statement}

\begin{figure*}
\begin{center}
\includegraphics[scale=0.5]{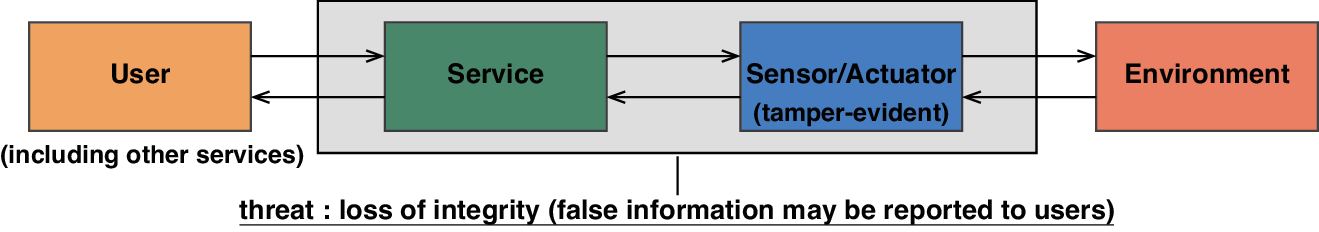}\\
\caption{Threat model for data integrity}\label{fig-problem}
\end{center}
\end{figure*}

The issue we address in this work is a generalization of the problem raised in our earlier study on RFID-based systems:
How can one ensure the authenticity---specifically the content, origin, time, and location---of data produced by sensors or actuators, even when the IoT services and networks relaying the data are not trustworthy?

As illustrated in Fig.~\ref{fig-problem}, a tamper-evident sensor or actuator interacts with the physical environment and reports data to a user via one or more intermediary services.
These services, which may include cloud-based IoT platforms, are potential points of compromise.
If any of these services are misconfigured, maliciously controlled, or otherwise untrustworthy, they may alter, drop, or delay the sensor data before it reaches the user---causing a loss of integrity.

Moreover, data may be intermittently lost due to hardware failures, network disruptions, or other systemic limitations.
Such loss should not invalidate the verifiability of surrounding data, nor should it hinder the detection of tampering attempts.
Therefore, a robust solution must not only ensure end-to-end authenticity but also tolerate occasional data loss without compromising trust in the remaining information.

Our goal is to design an architecture that satisfies these requirements by minimizing trust assumptions on intermediate systems, and by enabling verifiable, tamper-evident logging and auditing of sensor data under real-world constraints.

\section{Proposed Method}

To address the problem of ensuring sensor data authenticity in untrusted environments, we propose a design that enables tamper-evident logging of sensor or actuator data and its verification via a blockchain-backed evidence service.
This method supports both sporadic and streaming data, while tolerating data losses without compromising verifiability.

\subsection{Overview}

\begin{figure*}
\begin{center}
\includegraphics[scale=0.5]{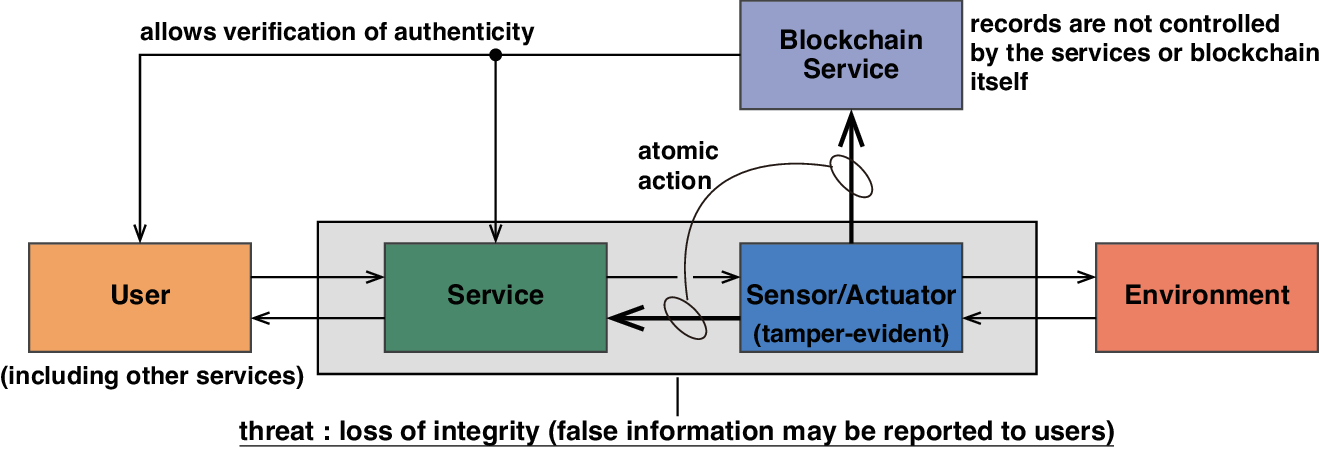}
{\footnotesize
\begin{itemize}
\item Sensor or actuator responds to requests from the service and also writes evidence to the blockchain service in a single atomic action, although it may fail.
\end{itemize}
}
\caption{System overview with blockchain logging}\label{fig-solution-overview}
\end{center}
\end{figure*}

As illustrated in Fig.~\ref{fig-solution-overview}, the proposed architecture introduces a blockchain service alongside the traditional user-service-sensor chain.
When a tamper-evident sensor or actuator receives a request from a service, it performs an atomic action: it responds to the request with a digitally signed readout, and simultaneously sends corresponding cryptographic evidence to the blockchain service \footnote{
As explained later, in the case of stream data, this atomic action occurs periodically, not every time.
}.
This atomicity ensures that evidence is available to verify what the service claims to have received---unless the entire action fails, in which case the absence of the evidence itself becomes observable, while, as shown in section~\ref{sec-handling-data-loss}, verifiability can be maintained in many cases.

The blockchain service is out of control of both the service provider and the sensor operator.
Therefore, it functions as a neutral third party to record the existence and integrity of data, rather than its contents.
Authenticity can be verified independently from any of the interacting parties.

The proposed architecture assumes that the sensor or actuator is tamper-evident, meaning that any unauthorized modification to the device or its outputs can be detected.
This assumption is crucial because our method verifies the integrity of the reported data, but not the behavior of the device itself.
Without tamper evidence, a compromised sensor might emit falsified values that are nevertheless cryptographically well-formed.
One promising approach to achieving tamper-evident behavior is through the use of Physical Unclonable Functions (PUFs) \cite{Herder2014:PUF}, which derive device-unique keys or signatures from intrinsic physical variations in the hardware.
These characteristics make PUFs difficult to clone or emulate, enabling the sensor to prove its identity and resist key extraction or spoofing.

\subsection{Data Structure}

\begin{figure*}
\begin{center}
\includegraphics[scale=0.5]{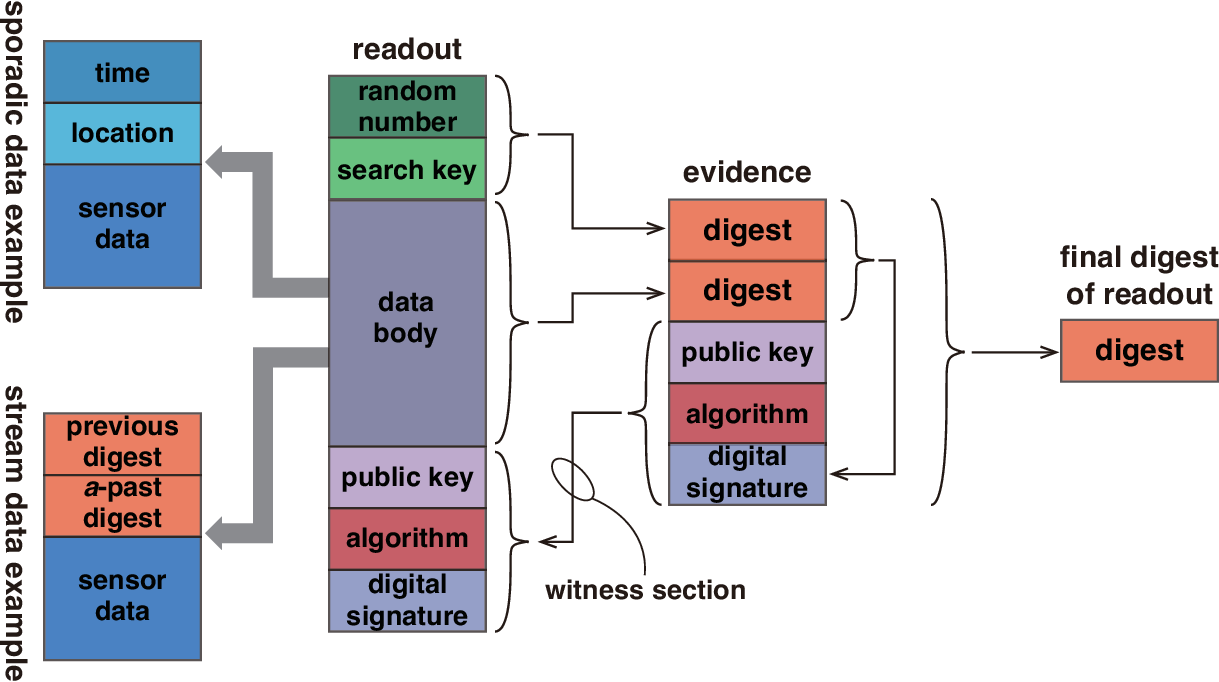}
{\footnotesize
\begin{itemize}
\item Public key identifies the sensor/actuator.
\item There can be multiple $<$random number, search key$>$ pairs, and data body can produce multiple digests for each section within.
These digests are collectively signed by the sensor/actuator.
\item For stream data, most of the time the witness section is omitted and the evidence is not sent to the blockchain service, unless it is a {\em checkpoint}, which is periodically inserted.
\end{itemize}
}
\caption{Sensor data structure and digests}\label{fig-solution-structure}
\end{center}
\end{figure*}

The structure of the data exchanged between the sensor and the evidence service is shown in Fig.~\ref{fig-solution-structure}.
Each {\em readout} includes a data body and metadata such as the time, location, or cryptographic linkage to previous outputs.
The readout is signed using the sensor's private key and is associated with a set of digests, forming the {\em witness section}.
These digests can be generated from different sections of the data, and optionally linked with a $<$random number, search key$>$ pair to anonymize traceability.
Furthermore, by structuring each readout into selectively verifiable segments, our approach allows partial disclosure of sensor data---balancing verifiability, privacy, and efficiency.

The {\em final digest} of the readout, incorporating the witness section with identifying information such as the public key and signature, is what gets committed to the blockchain (via Merkle aggregation within the blockchain service).
For stream data, the witness section is often omitted to reduce computational load and bandwidth, except for periodically inserted {\em checkpoints} that ensure continuity of verifiability.

\subsection{Handling Data Loss}\label{sec-handling-data-loss}

\begin{figure*}
\begin{center}
\includegraphics[scale=0.5]{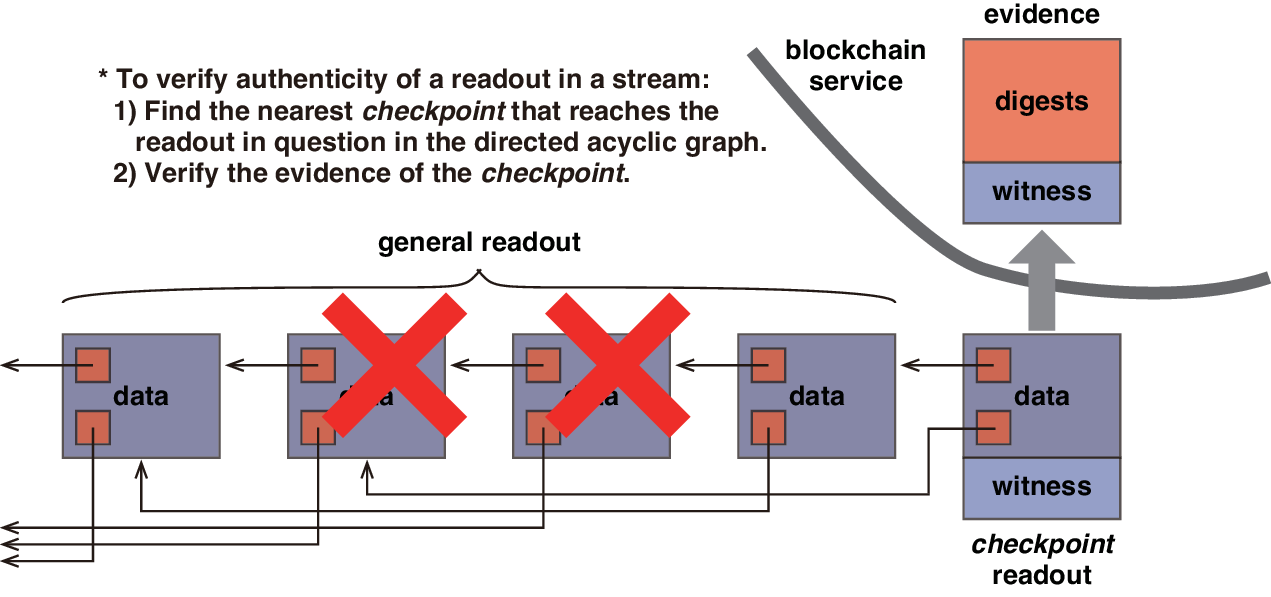}
{\footnotesize
\begin{itemize}
\item Each data stores the $a$-past digest, where $a = 3$ in this case, as well as the previous digest.
\item This tolerates up to $a - 1$ consecutive data losses.
\item If a {\em checkpoint} or all it points are lost, or if the atomic action to store the evidence fails, we just have to wait for another {\em checkpoint} to have recent data verifiable.
\end{itemize}
}
\caption{Redundant digest chaining under data loss}\label{fig-solution-data-loss}
\end{center}
\end{figure*}

To tolerate intermittent data loss, particularly in stream scenarios, our method employs a hash-chained digest structure adopted from previous relevant study \cite{Golle2001:Stream} as shown in Fig.~\ref{fig-solution-data-loss}.
Each data point records not only the previous digest, but also an additional $a$-past digest (e.g., three steps back in case $a = 3$).
This redundancy creates a directed acyclic graph (DAG) of digests, allowing recovery of verifiability even if up to $a - 1$ consecutive readouts are lost.

When verifying the authenticity of a particular readout, one can find the nearest reachable checkpoint and validate its evidence stored in the blockchain.
As long as some checkpoints are intact, recent data can be re-linked and verified without requiring full data retention.

This mechanism enables our method to remain robust against partial transmission failures while still preserving the end-to-end verifiability of the data stream.

\section{Simulation and Evaluation}

To assess the effectiveness of our method in terms of tolerance to data loss, we performed simulations using a synthetic stream of 10,000 readouts.
Each readout was randomly marked as lost with a configurable loss probability $p \in [0, 0.5]$.
A signed checkpoint was inserted every $s$ readouts, and each readout included digests linking it to both the immediately preceding readout and to the $a$-past readout (i.e., $a$-step back).
A readout was considered verifiable if it could be reached from at least one checkpoint via a valid chain of available digests.

The results were plotted in terms of the percentage of verifiable readouts as a function of $p$, for different configurations of 
$s$ and $a$.
The results were obtained from single simulation runs, but the observed trends remained consistent across multiple repeated trials.

\subsection{Effect of Signature Interval $s$}

\begin{figure}
\begin{center}
\includegraphics[scale=0.5]{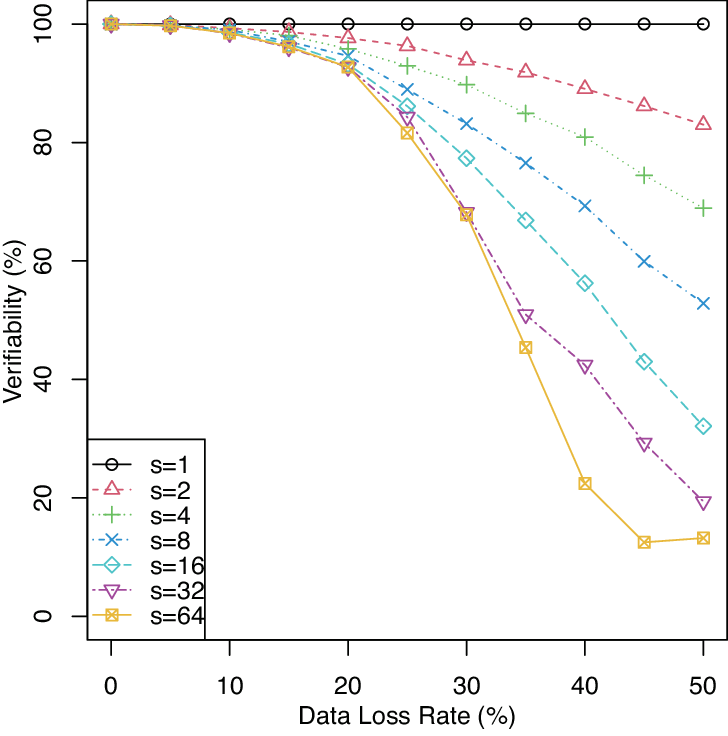}\\
\caption{Effect of signature interval $s$ ($a=10$)}\label{fig-eval-s}
\end{center}
\end{figure}

As shown in Fig.~\ref{fig-eval-s}, decreasing the signature interval $s$ (i.e., increasing the frequency of checkpoints) significantly improves the overall verifiability.
When every readout is a checkpoint ($s=1$), all non-lost readouts are trivially verifiable.
As $s$ increases, a single run of consecutive losses can disconnect larger blocks of readouts from their nearest checkpoints, degrading verifiability.

This result highlights a key design trade-off between signature overhead and verification robustness.

\subsection{Effect of a-past Linkage $a$}

\begin{figure}
\begin{center}
\includegraphics[scale=0.5]{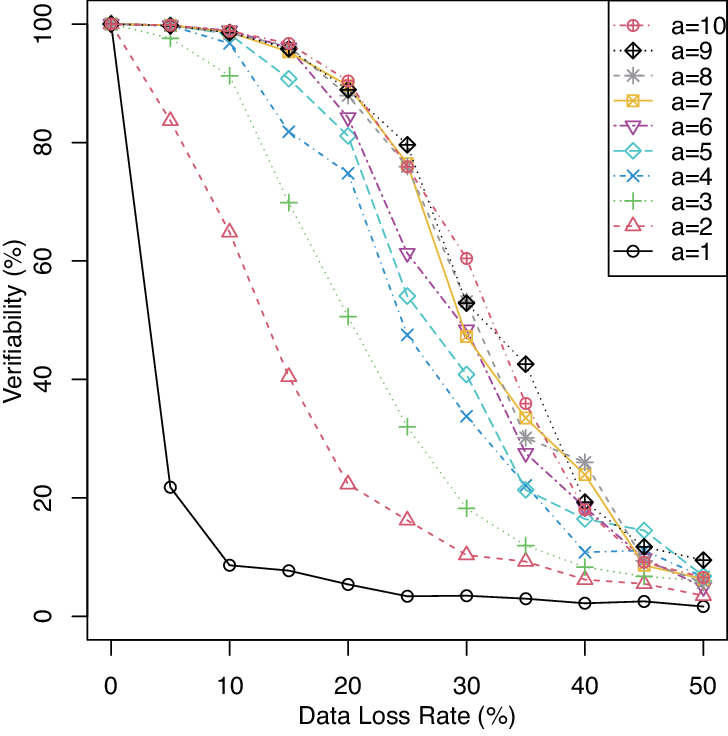}\\
\caption{Effect of $a$-past Linkage ($s=100$)}\label{fig-eval-a}
\end{center}
\end{figure}

Fig.~\ref{fig-eval-a} shows how the offset for the redundant link (parameter $a$) affects resilience to loss when the signature interval 
$s$ is fixed at 100.
With only one-step backward link ($a=1$), the structure behaves exactly like a singly linked list: one loss in the chain breaks all subsequent verifiability.
However, increasing $a$ dramatically improves fault tolerance.
For $a=3$, for example, verification is still possible even if up to two consecutive readouts are lost, since multiple paths exist to reach a checkpoint.

\subsection{Saturation Behavior for Large $a$}

\begin{figure}
\begin{center}
\includegraphics[scale=0.5]{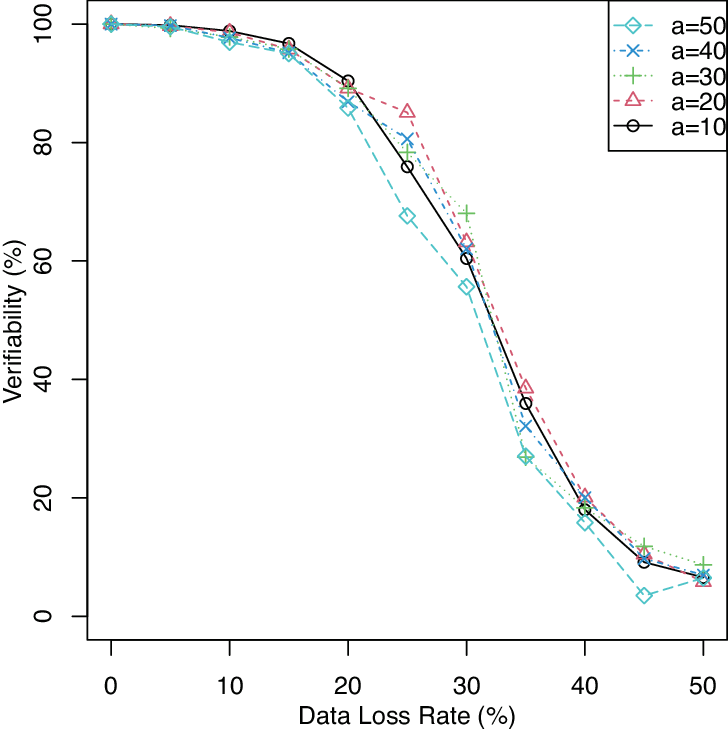}\\
\caption{Saturation effect with large $a$ ($s=100$)}\label{fig-eval-large-a}
\end{center}
\end{figure}

Fig.~\ref{fig-eval-large-a} extends the analysis to larger values of $a$.
We observe diminishing returns beyond around $a=10$.
This suggests a saturation effect, where increasing redundancy further provides little additional benefit.

\subsection{Sample Implementation}

To further validate the feasibility of our proposed architecture, we implemented a prototype system that records and verifies readouts under varying loss conditions, available at GitHub\footnote{{\scriptsize https://github.com/beyond-blockchain/bbc2-examples/tree/develop/file-recorder}}.
The sample tests demonstrate that the proposed digest-chaining and checkpointing mechanism remains effective under realistic parameter settings, and the results were compatible with the trends observed in our simulations.

\section{Discussion}

Our simulations confirm that the use of $a$-past digest links substantially enhances the verifiability of sensor data under conditions of partial loss or recording failure.
Originally motivated by applications involving streamed data---such as surveillance cameras or continuous monitoring systems---the $a$-past linkage design also proves valuable in more sporadic but non-trivially structured contexts, such as logistics applications.

In such settings, data may not arrive as a fully continuous stream, but still exhibits non-isolated patterns of loss.
For instance, RFID readers in a logistics chain may experience temporary failures or connectivity issues, resulting in blocks of missed readouts.
By equipping each readout with a redundant backward link---i.e., to the previous record and to an earlier one---it becomes possible to maintain the integrity of data verification even when some readouts are lost or an atomic logging action fails.

Importantly, our evaluation shows that moderate redundancy (e.g., 3 to 10-step backward links) is sufficient to cover a wide range of loss patterns without imposing excessive computational or transmission overhead.
This suggests that the $a$-past mechanism is not only applicable to high-rate streaming scenarios, but also highly beneficial in semi-structured, event-driven, or industrial IoT environments where loss is correlated, yet bounded.

In practical deployments, these findings support the use of digest-linked structures as a lightweight yet resilient means of achieving end-to-end verifiability, even in the absence of trusted intermediate services.
By carefully configuring the redundancy parameters according to the expected loss profile and device capabilities, system designers can achieve a favorable balance between robustness and cost.

Moreover, it is not necessary for parameters such as $a$ (the offset for the redundant links) or $s$ (the checkpoint interval) to remain fixed across all situations.
These values can be dynamically adjusted in response to observed loss rates, system conditions, or application requirements.
For example, a device experiencing intermittent network instability might temporarily increase $a$ to improve resilience (to tolerate up to $a-1$ consecutive loss of data), while reducing it when operating under stable conditions.
Such adaptive configurations open the door to self-optimizing verification frameworks that balance robustness and efficiency in real time.

Beyond industrial and logistics applications, the proposed method holds promise for humanitarian technology use cases.
In disaster-stricken or low-infrastructure regions, where sensors may intermittently fail or lose connectivity, maintaining verifiability of available data can support rapid damage assessment, supply tracking, or public health monitoring.
By enabling end-to-end trust without reliance on centralized verification services, our approach can contribute to building resilient, decentralized sensing systems in critical and resource-constrained environments.

\section{Related Work}

Recent studies have explored blockchain-based methods for securing IoT data and improving trust in sensor-originated information.
\cite{Li2019:IoT} proposed two signature schemes, Dynamic Tree Chaining (DTC) and Geometric Star Chaining (GSC), which amortize digital signatures across structured blocks to enable verifiable partial data retrieval.
These schemes are particularly efficient for sampled data and offer lightweight authentication suitable for resource-constrained IoT devices.
However, they do not address unintentional data loss during transmission.
In particular, DTC is sensitive to missing events due to its Merkle-tree structure, while GSC assumes that lost samples are a result of controlled sampling policies rather than network failures or logging errors.

\cite{Lee2025:IoT} developed a Blockchain-based Mobile IoT System (BMIS) that employs a dual-path architecture: sensor data are sent both to a cloud service (ThingSpeak) for real-time visualization and to a public blockchain for immutable storage.
Their design demonstrates high reliability in controlled conditions but relies on uninterrupted transmission to preserve data integrity.
Data loss is treated as an experimental limitation (e.g., a 98.35\% upload success rate), and no structural mechanism is included to detect or recover from lost readouts.

\cite{GomezRivera2022:IoT} introduced BloSPAI, a protocol for SCADA systems that combines SRAM-based Physical Unclonable Functions (PUFs) with a permissioned blockchain to continuously authenticate field sensors.
While BloSPAI strengthens hardware-level trust and provides robust device authentication, it assumes reliable communication and does not explicitly address message loss or delayed arrival.
The system also requires secure enrollment and stable environmental conditions to maintain the reliability of the PUF responses.

Our approach addresses not only data authenticity and integrity, but also verifiability under data loss.
Using a digest chain where each readout links to its predecessor and another prior readout (e.g., an $a$-past one), we form a sparse DAG that allows reconstructing verification paths even with missing data.
This design ensures robustness against bursty or unpredictable losses, making it suitable for logistics and intermittent sensing scenarios.

Compared to DTC or GSC, which rely on structured sampling or degrade under missing data, our method remains verifiable without such assumptions.
Unlike BloSPAI, which requires secure enrollment for device authentication, we rely on lightweight self-managed public keys without centralized provisioning.

\section{Conclusions}

We have presented a lightweight and verifiable logging method for IoT sensor and actuator data that remains robust even when intermediate services are untrusted and data losses occur.
The proposed architecture enables each sensor to perform an atomic operation that both transmits its signed readout to the service and submits cryptographic evidence to an independent blockchain-based service.
By organizing this evidence using digest chains and Merkle roots, users can later verify the authenticity of data with minimal trust assumptions.

A key innovation in our design is the use of a configurable-offset $a$-past linkage adopted from \cite{Golle2001:Stream}: each readout points not only to its immediate predecessor, but also to a specific earlier readout $a$ steps back.
This structure enhances resilience to data loss, particularly in the presence of bursty or clustered failures, without requiring large numbers of redundant links.
Our simulation results demonstrate that even modest values of $a$ (e.g., 3 to 10) significantly improve verifiability across a wide range of data loss conditions and checkpoint intervals.

While originally motivated by streaming applications, the $a$-past linkage mechanism also benefits semi-structured or event-driven data sources such as those found in logistics systems.
In these contexts, data may be sporadic yet still vulnerable to correlated failures or atomic action disruptions.
Applying $a$-past linking in such scenarios can thus provide substantial gains in reliability without imposing continuous logging overhead.

In practice, the parameters $a$ and $s$ need not be fixed.
They may be adapted dynamically according to device behavior, network conditions, or application requirements.
This flexibility offers promising directions for the development of adaptive and context-aware verifiability frameworks in future IoT deployments.

\section*{Acknowledgment}
We thank Prof.~Hiroshi Watanabe of National Yang Ming Chiao Tung University, and Tsuneo Kato and Shusaku Torisawa of Beyond Blockchain Inc. for their valuable discussions.
We also thank 
LAPIS Technology Co., Ltd. for their knowledge of sensors.

\bibliographystyle{plain}
\bibliography{general-iot-r10-htc}

\end{document}